\documentstyle[12pt]{article}
\input epsf.sty
\def\ga{\mathrel{\raise.3ex\hbox{$>$\kern-.75em\lower1ex\hbox{$\sim$}}}}
\def\la{\mathrel{\raise.3ex\hbox{$<$\kern-.75em\lower1ex\hbox{$\sim$}}}}
\def\he#1{\hbox{${}^{#1}$He}}
\def\li#1{\hbox{${}^{#1}$Li}}
\def\beq{\begin{equation}}
\def\eeq{\end{equation}}

\begin{document}
\pagestyle{plain}
\baselineskip=13pt
\rightline{UMN--TH--1427/96}
\rightline{astro-ph/9605068}
\rightline{May 1996}
\vspace*{1.2cm}
\begin{center}
{\large{
WHY DO WE NEED NON-BARYONIC DARK MATTER? 
\footnote{To be published in the proceedings of the XXXIst 
Recontres de Moriond,
Les Arcs, France, January 20-27 1996.}
 }}
\end{center}
~\newline

\baselineskip=2ex
\begin{center}
{\large Keith A.~Olive
}\\
{\large \it
{School of Physics and Astronomy,
University of Minnesota,\\ Minneapolis, MN 55455, USA}}

\vspace*{3cm}
{\bf Abstract}
\end{center}
Observational evidence along with theoretical arguments which call for 
non-baryonic dark matter are reviewed. A brief summary of the dark 
matter session
is included.

\vspace*{4.5ex}
\baselineskip=3ex
 There is increasing evidence  
that relative to the visible matter in the Universe, which 
is in the form of baryons, there is considerably more
 matter in the Universe 
that we don't see \cite{dm}.
	Here, I will review some of the motivations for dark
 matter in the Universe.  The best observational evidence 
is found on the scale of galactic halos and comes from the 
observed flat rotation curves of galaxies. There is also mounting
evidence for dark matter in elliptical galaxies as well as clusters
of galaxies coming from X-ray observations of these objects.
Also, direct evidence has been obtained through the study of
gravitational lenses.
In theory, we 
believe there is much more matter because 1) inflation 
tells us so (and there is at present no good alternative to inflation)
 and 2) our current understanding of galaxy 
formation only makes sense if there is more matter than we see.  
	One can also make a strong case for the existence of non-baryonic 
dark matter in
particular. The recurrent problem  with baryonic dark matter
is that not only is it very difficult to hide baryons, but 
the standard model of primordial nucleosynthesis
would have to discarded
if all of the dark matter is baryonic. 
 Fortunately, as will be covered at length
in these proceedings, there are several attractive alternatives to baryonic
dark matter.

Before embarking on the subject of dark matter, it will be useful to review 
the relevant quantities from the standard big bang model.
In a Friedmann-Robertson-Walker Universe, the expansion rate of the 
Universe (the
Hubble parameter) is related to the energy density $\rho$ and 
curvature constant
$k$ by
\beq
	H^2 = \left({\dot{R} \over R}\right)^2 = {8 \pi G \over 3} \rho -
  {k \over R^2}
\label{H}
\eeq
assuming no cosmological constant, where $k = \pm 1, 0$ for a closed, open
or spatially flat Universe, and $R$ is the cosmological scale factor.
When $k = 0$, the energy density takes its ``critical" value,
$\rho = \rho_c = 3 H^2/8\pi G = 1.88 \times 10^{-29}  h_o^2$ g cm$^{-3}$
where $h_o = H_o /100$ km s$^{-1}$ Mpc$^{-1}$ is the scaled present
value of the Hubble parameter. The cosmological density parameter is defined
by $\Omega = \rho/\rho_c$ and
by rewriting eq. (\ref{H}) we can relate $k$ to $\Omega$ and $H$ by
\beq
	{ k \over R^2 } = (\Omega - 1) H^2 
\label{o1}		
\eeq
so that $k = +1, -1, 0$ corresponds to $\Omega > 1, < 1, = 1$. 

In very broad terms, observational limits on the cosmological parameters
are: $0.2 \la \Omega \la 2$ and $0.4 \la h_o \la 1.0$ \cite{tonry}.
The cosmological density is however sensitive to the
particular scale being observed (at least on small scales).
Accountably visible matter contributes in total only a small fraction
to the overall density, giving $\Omega_V \sim .003 -.01$.  
In the bright central parts of galaxies, the density is larger
 $\Omega  \sim
0.02- 0.1$.  On larger scales, that of
 binaries and small groups of galaxies, 
 $\Omega   \simeq  0.05 -0.3$.  On even larger scales the density
may be large enough to support $\Omega \simeq
1.0$.  Though there are no astronomical observations to 
support $\Omega > 1$, 
limits based on the deceleration of the Universe only 
indicate \cite{tonry} that $\Omega \la 2$.

The age of the Universe is also very sensitive to these parameters. 
Again, in the absence of a cosmological constant we have,
\beq
H_o t_U = \int^1_0 (1 - \Omega + \Omega/x)^{-1/2} dx
\eeq
For $t_U > 13$ Gyr, $\Omega h_o^2 < 0.25$ if $h_o > 0.5$ and
$\Omega h_o^2 < 0.45$ if $h_o > 0.4$.  While for 
$t_U > 10$ Gyr, $\Omega h_o^2 < 0.8$ if $h_o > 0.5$ and
$\Omega h_o^2 < 1.1$ if $h_o > 0.4$.

\medskip

There is, in fact, good evidence for dark matter 
on the scale of galaxies (and their halos). 
 Assuming that galaxies are in virial equilibrium,
 one expects that by Newton's Laws one can relate 
the mass at a given distance $r$, from the center of 
a galaxy to its rotational velocity
\beq
	M(r) \propto v^2 r/G_N 	
\eeq
The rotational velocity, $v$, is measured \cite{fg,bos,rft}
 by observing 21 cm 
emission lines in HI regions (neutral hydrogen) beyond the point 
where most of the light in the galaxy ceases.  A compilation 
of nearly 1000
 rotation curves of spiral galaxies have been plotted in 
\cite{pss} as a function
of $r$ for varying brightnesses.  If the bulk of the mass is 
associated with light, then beyond the point where most of the light 
stops $M$ would be 
constant and $v^2  \propto 1/r$.  This is not the case, as
the rotation
 curves appear to be flat, i.e., $v \sim$ constant outside the
 core of the galaxy. This implies that $M \propto r$ beyond the point
 where the light stops.  This is one of the strongest pieces of 
evidence for the existence of dark matter. Velocity measurements indicate
dark matter in elliptical galaxies as well \cite{sag}.

Galactic rotation curves are not the only observational indication for the
existence of dark matter.  X-ray emitting hot gas in elliptical galaxies 
also provides an important piece of evidence for dark matter. 
 As an example, 
consider the large elliptical M87.  The detailed profiles of the temperature
and density of the hot X-ray emitting gas have been mapped out \cite{fgo}.
By assuming hydrostatic equilibrium, these measurements allow 
one to determine 
the overall mass distribution in the galaxy necessary to bind the hot gas.
Based on an isothermal model with temperature $kT = 3$keV (which leads
to a conservative estimate of the total mass), Fabricant and
 Gorenstein \cite{fgo}
predicted that the total mass out to a radial distance
 of 392 Kpc, is $5.7 \times 10^{13} M_\odot$
whereas the mass in the hot gas is only $2.8 \times 10^{12} 
M_\odot$ or only 5\%
of the total. The visible mass is expected to contribute 
only 1\% of the total.
The inferred value of $\Omega$ based on M87 would be $\sim 0.2$.

M87 is not the only example of an elliptical galaxy in which X-ray emitting
hot gas is observed to indicate the presence of dark matter.  
At this meeting,
Foreman \cite{F1}, showed several examples of ellipiticals 
with large mass to
light ratios.  For example in the case of N4472, while the 
optical observations
go out to 25 kpc, the X-ray gas is seen out to 75 kpc, 
indicating M/L's of about
60 at 70 kpc and up to 90 at 100 kpc. Similar inferences regarding the
existence of dark matter can be made from the X-ray emission 
from small groups of
galaxies \cite{mush,F2}.

On very large scales, it is possible to get an estimate of $\Omega$ from the
distribution of peculiar velocities. On scales, $\lambda$, 
where perturbations, $\delta$, are still
small, peculiar velocities can be expressed \cite{peeb}
 as $v \sim H \lambda \delta \Omega^{0.6}$.
 On these scales, measurements of the peculiar velocity field from 
the IRAS galaxy catalogue indicate that indeed $\Omega$ 
is close to unity \cite{iras}.
Another piece of evidence on large scales, is available from 
gravitational lensing
\cite{tyson}. The systematic lensing of the roughly 150,000 
galaxies per deg$^2$
at redshifts between $z = 1 - 3$ into arcs and arclets allow 
one to trace the
matter distribution in a foreground cluster. Van Waerbeke 
discussed recent results
of weak gravitational lensing looking at systems, which if 
virialized, have mass to
light ratios in the range 400--1000 and correspond to values 
of $\Omega$ between
0.25 and 0.6 \cite{VW}.
 Unfortunately, none of these 
observations reveal the identity of the dark matter.

Theoretically, there is no lack of support for the dark matter hypothesis.
The standard big bang model including inflation almost requires that
$\Omega = 1$ \cite{oinf}.
The simple and
unfortunate fact that at present we do not even know whether $\Omega$ is
larger or smaller than one indicates that we do not know the sign of
the curvature term, further implying that it is subdominant in Eq.
(\ref{H})
\beq
	  { k \over R^2 } < {8 \pi G \over 3} \rho
\eeq
In an adiabatically expanding Universe, $R \sim T^{-1}$   where $T$ is the
temperature of the thermal photon background.  Therefore the quantity
\beq
	\hat{k} = { k \over R^2 T^2} <  
{8 \pi G \over 3 T_o^2} < 2 \times 10^{-58}
\label{khat}
\eeq
is dimensionless and constant in the standard model.  The smallness of
$\hat{k}$ is known as
the curvature problem and can be resolved by a period of inflation.
Before inflation, let us write $R = R_i$, $T = T_i$  and $R \sim T^{-1}$. 
 During
inflation, $R \sim T^{-1} \sim e^{Ht}$, where $H$ is constant.  
After inflation, $R =
R_f  \gg R_i$  but $T = T_f  = T_R  \la T_i$  where $T_R$  is the temperature
 to which the
Universe reheats.  Thus $R \not\sim T^{-1}$   and $\hat{k} \rightarrow 0$
 is not constant.  But from
Eqs. (\ref{o1}) and (\ref{khat}) if $\hat{k} \rightarrow 0$ then
 $\Omega \rightarrow 1$, and since typical
inflationary models contain much more expansion than is necessary, $\Omega$
becomes exponentially close to one.

	If this is the case and $\Omega = 1$, then we know two things:  Dark
matter exists, since we don't see $\Omega = 1$ in luminous objects, and most
(at least 90\%) of the dark matter is not baryonic.  The latter
conclusion is a result from big bang nucleosynthesis \cite{wssok,fo}, which
constrains the baryon-to-photon ratio $\eta = n_B/n_\gamma$  to
\beq
	1.4 \times 10^{-10} < \eta < 3.8 \times 10^{-10}   		
\eeq
which corresponds to a limit on $\Omega_B$
\beq
	0.005 < \Omega_B < 0.09		
\eeq
for $0.4 \la h_o \la 1.0$.  Thus $1-\Omega_B$ is not only dark but
also non-baryonic. I will return to big bang nucleosynthesis below.

	Another important piece of theoretical evidence for dark
 matter comes from the simple fact that we are here living in a galaxy.
The type of perturbations produced
 by inflation \cite{press} are, in most models,
 adiabatic perturbations ($\delta\rho/\rho \propto
 \delta T/T)$, and I
 will restrict my attention to these.  Indeed, the perturbations
produced by inflation
 also have the very nearly scale-free spectrum described by 
Harrison and Zeldovich \cite{hz}.  When produced, scale-free perturbations 
fall off as $\frac{\delta \rho}{\rho} \propto l^{-2}$
 (increase as the square of the wave number). 
 At early times $\delta\rho/\rho$ grows as $t$
 until the time when the horizon scale (which is
 proportional to the age of the Universe) is comparable to $l$.  At later 
times, the growth halts (the mass contained within the volume $l^3$  
has become smaller than the Jean's mass) and   
  $\frac{\delta \rho}{\rho} = \delta$ (roughly) independent of the scale $l$.
 When the Universe becomes matter dominated, the Jean's mass 
drops dramatically and growth continues as $\frac{\delta \rho}{\rho} \propto
 R \sim 1/T$.  For an overview of the evolution of density perturbations  and
the resulting observable spectrum see \cite{lr}. The transition to matter 
dominance
is determined by setting the energy densities in radiation
(photons and any massless  neutrinos) equal to the energy density in  matter
(baryons and any dark matter).  For three massless  neutrinos and baryons (no
dark matter), matter dominance begins at
\beq
	T_m  = 0.22 m_B \eta	
\eeq
and for $\eta < 3.8 \times 10^{-10}$, this corresponds to
$T_m < 0.08$ eV.

The subsequent non-linear growth in $\delta \rho/\rho$ was discussed in these
sessions in some detail at this meeting. In particular, there was
 a considerable
discussion of the effects of the non-linear regime on the power 
spectrum and the
appearance of non-Gaussian features such as skewness and kurtosis \cite{bern}. 
Colombi
\cite{col} discussed the 2- and 3-point correlation functions. Numerous
simulations were presented to reflect the 
hydrodynamics of galaxy formation  \cite{tor} and the origin of the
 large scale
bias \cite{kauf}.

	Because we are considering adiabatic perturbations,
 there will be anisotropies produced in the microwave 
background radiation on the order of $\delta T/T \sim \delta$.  
The value of $\delta$, the amplitude of the density fluctuations at horizon 
crossing, has now been determined by COBE \cite{cobe}, $\delta =
(5.7 \pm 0.4) \times 10^{-6}$.  Without the existence of dark matter,
 $\delta \rho/\rho$ in baryons could then achieve a maximum value of only
$\delta\rho/\rho \sim A_\lambda \delta(T_m/T_o)  
\la 2 \times 10^{-3}A_\lambda$,
where $T_o = 2.35 \times 10^{-4}$ eV is the present temperature of the
microwave background and $A_\lambda
\sim 1-10$ is a scale dependent growth factor. 
 The overall growth in $\delta \rho / \rho$ is too small to argue
 that growth has entered a nonlinear regime needed to explain
 the large value ($10^5$) of $\delta\rho/\rho$ in galaxies.

	Dark matter easily remedies this dilemma in the following way.
 The transition to matter dominance is determined by setting equal 
to each other the energy densities in radiation (photons and any massless 
neutrinos) and matter (baryons and any dark matter). 
While if we suppose that there exists 
dark matter with an abundance $Y_\chi = n_\chi/n_\gamma$  
(the ratio of the number density of $\chi$'s to photons) then
\beq
	T_m  = 0.22 m_\chi Y_\chi	
\eeq
Since we can write $m_\chi Y_\chi/{\rm GeV} = \Omega_\chi
 h^2/(4 \times 10^7)$,
we have $T_m/T_o = 2.4 \times 10^4 \Omega_\chi h^2$ which is 
to be compared with
350 in the case of baryons alone.  
The baryons, although still bound to the radiation until 
decoupling,  now see deep potential wells formed by the dark matter
 perturbations to fall into and are no longer required to 
grow at the rate $\delta \rho/\rho \propto R$.

	With regard to dark matter and galaxy formation, all forms 
of dark matter are not equal.  They can be distinguished 
by their relative temperature at $T_m$ \cite{bond}. Particles which are still 
largely relativistic at $T_m$ (like neutrinos or other particles with 
$m_\chi < 100$ eV) have the property \cite{free} that 
(due to free streaming) they
erase perturbations 
 out to very large scales given by the Jean's mass
\beq
	M_J  = 3 \times 10^{18}  {M_\odot \over {m_\nu}^2(eV)}	
\label{mj}
\eeq
Thus, very large scale structures form first and galaxies
 are expected to fragment out later.  Particles with this 
property are termed hot dark matter particles.  
Cold particles ($m_\chi > 1$ MeV) have the opposite behavior. 
 Small scale structure forms first aggregating to form 
larger structures later.  Neither candidate is completely 
satisfactory when the resulting structure is compared to the 
observations.  For more details, I refer the reader to reviews 
in refs. \cite{dm}.

The most promising possibility we have for unscrambling the various possible
scenarios for structure formation is the careful analysis 
of the observed power
spectrum. Rewriting the density contrast in Fourier space, 
\beq
\delta(k) \propto \int d^3x {\delta \rho \over \rho}(x) e^{ik \cdot x}
\eeq
the power spectrum is just
\beq
P(k) = \langle |\delta(k)|^2 \rangle
\eeq
which is often written in term of a transfer function and a power law, $P \sim
T(k) k^n$. (The flat spectrum produced by inflation has $n=1$.)  The data
contributing to
$P(k)$ include observations of galaxy distributions, velocity 
distributions and of
course the CMB
\cite{vog}.  However to make a comparison with our theoretical 
expectations, we
rely on n-body simulations and a deconvolution of the theory 
from the observations.
Overall, there is actually a considerable amount of concordance with our
expectations. The velocity distributions indicate that 
$0.3 < \Omega < 1$ and the
power spectrum is roughly consistent with an $\Omega = 1$, 
and $n=1$, cold dark
matter Universe.

Future surveys \cite{fut,mand} will however, most certainly 
dramatically improve
our understanding of the detailed features of the power spectrum and their
theoretical interpretations.  We should in principle be able to 
distinguish between
a mixed dark matter and a cold dark matter $\Omega = 1$ Universe, 
whether or not
$\Omega < 1$, with CDM, the presence of a cosmological constant, 
or a tilt in the
spectrum. Zurek \cite{zur}, stressed the importance of numerical 
simulations in
this context. Future probes of the small scale anisotropy \cite{mand,del}
 should in addition be able to determine the values of not only $\Omega$,
but also $\Omega_B$ and $h_o$ to a high degree of accuracy 
through a careful analysis of the Doppler peak in the power spectrum.

\medskip

	Accepting the dark matter hypothesis, the first choice for a
 candidate should be something we know to exist, baryons.  
Though baryonic dark matter can not be the whole story if $\Omega = 1$, 
 the identity of the
 dark matter in galactic halos, which appear to contribute at the 
level of $\Omega \sim 0.05$,  remains an important question needing to be
resolved.  A baryon density of this magnitude is not excluded by 
nucleosynthesis. 
 Indeed we know some of the baryons are dark since $\Omega \la 0.01$ 
in the disk of the galaxy.

It is quite difficult, however, to hide large amounts 
of baryonic matter \cite{hio12}. Sites for halo baryons that 
have been discussed
include snowballs, which tend to sublimate, cold hydrogen gas, which 
must be supported against collapse, and hot gas, which can be excluded by the 
X-ray background. Stellar objects (collectively called MACHOs for macroscopic
compact halo objects) must either be so
small ( M $< 0.08$ M$_\odot$) so as not to have begun nuclear burning or so
massive so as to have undergone total gravitational collapse without the 
ejection of heavy elements.  Intermediate mass stars are generally quite
problematic because either they are expected to still reside on the 
main-sequence
today and hence would be visible, or they would have produced an 
excess of heavy
elements.

On the other hand, MACHOs are a candidate which are testable by the
gravitational microlensing of
stars in a neighboring galaxy such as the LMC \cite{pac}. By observing
millions of stars and examining their intensity as a function of time,
it may be possible to determine the presence of dark objects in our halo.
It is expected that during a lensing event, a star's  intensity will rise
 in an achromatic fashion over  a period
$\delta t
\sim 3$ $\sqrt{M/.001 M_\odot}$ days.
Indeed, microlensing candidates have been found \cite{macho}. For low mass
objects, those with $M < 0.1M_\odot$, it appears however that 
the halo fraction
of Machos is small, $\approx 0.19^{+.16}_{-.10}$ \cite{m1}. 
There have been some
recent reports of events with longer duration \cite{m2} leading to the
speculation of a white dwarf population in the halo. Though it is too early to
determine the implications of these observations, they are very encouraging in
that perhaps this issue can and will be decided.

\medskip

The degree to which baryons can contribute to dark matter depends
ultimately on the overall baryon contribution to $\Omega$ which is 
constrained by
nucleosynthesis. Because of its importance regarding the issue of 
dark matter and
in particular non-baryonic dark matter, I want to review the status of big
bang nucleosynthesis.

 Conditions for the synthesis of the light elements were
attained in the early Universe at temperatures  $T \la $ 1 MeV, 
corresponding to
an age of about 1 second.  Given a single input parameter, the 
baryon-to-photon
ratio,
$\eta$, the theory is capable of predicting the primordial abundances of
D/H, \he3/H, \he4/H and \li7/H. The comparison of these predictions with
the observational determination of the abundances of the light elements
not only tests the theory but also fixes the value of $\eta$.

The dominant product of big bang nucleosynthesis is \he4 resulting in an
abundance of close to 25 \% by mass. In the standard model,
 the \he4 mass fraction
depends only weakly on
$\eta$. When we go beyond the standard model, the
\he4 abundance is very sensitive to changes in the expansion rate which 
can be related to the effective number of neutrino flavors.
 Lesser amounts of the other light elements are produced:
D and \he3 at the level of about $10^{-5}$ by number, and \li7 at the level of
$10^{-10}$ by number.

There is now a good collection of abundance information on the \he4 mass
fraction, $Y$, O/H, and N/H in over 50 extragalactic HII 
(ionized hydrogen) regions
\cite{p,evan,iz}. The observation of the heavy elements is 
important as the helium
mass fraction observed in these HII regions has been augmented by some stellar
processing, the degree to which is given by the oxygen and nitrogen abundances.
In an extensive study based on the data in \cite{p,evan}, it was found
\cite{osa}   that the data is well represented by a linear correlation for
$Y$ vs. O/H and Y vs. N/H. It is then expected that the primordial abundance
of \he4 can be determined from the intercept of that relation.  
The overall result
of that analysis indicated a primordial mass fraction, 
 $Y_p  = 0.232 \pm 0.003$.
In \cite{osc}, the stability of this fit was verified by a 
Monte-Carlo analysis
showing that the fits were not overly sensitive to any particular HII region.
In addition, the data from \cite{iz} was also included, yielding a \he4 mass
fraction \cite{osc}
\beq
Y_p = 0.234 \pm 0.003 \pm 0.005
\label{he4}
\eeq
The second uncertainty is an estimate of the systematic uncertainty in the
abundance determination. 

The \li7 abundance
is also reasonably well known.
 In old,
hot, population-II stars, \li7 is found to have a very
nearly  uniform abundance. For
stars with a surface temperature $T > 5500$~K
and a metallicity less than about
1/20th solar (so that effects such as stellar convection may not be important),
the  abundances show little or no dispersion beyond that which is
consistent with the errors of individual measurements.
Indeed, as detailed in \cite{sp}, much of the work concerning
\li7 has to do with the presence or absence of dispersion and whether
or not there is in fact some tiny slope to a [Li] = $\log$ \li7/H + 12 vs.
T or [Li] vs. [Fe/H] relationship.
There is \li7 data from nearly 100 halo stars, from a 
 variety of sources. I will use the value given in \cite{mol} 
as the best estimate
for the mean \li7 abundance and its statistical uncertainty in halo stars 
\beq
{\rm Li/H = (1.6 \pm 0.1 {}^{+0.4}_{-0.3} {}^{+1.6}_{-0.5}) \times 10^{-10}}
\label{li}
\eeq
 The first error is statistical, and the second
is a systematic uncertainty that covers the range of abundances
derived by various methods. 
The third set of errors in Eq. (\ref{li}) accounts for
 the possibility that as much as half
of the primordial \li7 has been
destroyed in stars, and that as much as 30\% of the observed \li7 may have been
produced in cosmic ray collisions rather than in the Big Bang.
 Observations of \li6,
Be, and B help constrain the degree to which these effects
play a role \cite{fossw}. For \li7, the uncertainties are clearly dominated by
systematic effects.

Turning to D/H, we have three basic types of abundance information:
1) ISM data, 2) solar system information, and perhaps 3) a primordial
abundance from quasar absorption systems.  The best measurement for ISM D/H
is \cite{linetal}
\beq
{\rm (D/H)_{ISM}} = 1.60\pm0.09{}^{+0.05}_{-0.10} \times 10^{-5}
\eeq
This value may not be universal
(or galactic as the case may be) since there may be some real dispersion 
of D/H in the ISM \cite{ferl}. The solar abundance of D/H is inferred from two
distinct measurements of \he3. The solar wind measurements of \he3 as well as 
the low temperature components of step-wise heating measurements of \he3 in
meteorites yield the presolar (D + \he3)/H ratio, as D was 
efficiently burned to
\he3 in the Sun's pre-main-sequence phase.  These measurements 
indicate that \cite{scostv,geiss}
\beq
{\rm \left({D +~^3He \over H} \right)_\odot = (4.1 \pm 0.6 \pm 1.4) \times
10^{-5}}
\eeq
 The high temperature components in meteorites are believed to yield the true
solar \he3/H ratio of \cite{scostv,geiss}
\beq
{\rm \left({~^3He \over H} \right)_\odot = (1.5 \pm 0.2 \pm 0.3) \times
10^{-5}}
\label{he3}
\eeq
The difference between these two abundances reveals the presolar D/H ratio,
giving,
\beq
{\rm (D/H)_{\odot}} \approx (2.6 \pm 0.6 \pm 1.4) \times 10^{-5}
\eeq
Finally, there have been several recent reported measurements of 
D/H is high redshift quasar absorption systems. Such measurements are in
principle capable of determining the primordial value for D/H and hence $\eta$,
because of the strong and monotonic dependence of D/H on $\eta$.
However, at present, detections of D/H  using quasar absorption systems
indicate both a 
high and  low value of D/H.  As such, we caution that these values may not
turn  out to represent the true primordial value.
The first of these measurements \cite{quas1} indicated a rather high D/H ratio,
D/H $\approx$ 1.9 -- 2.5 $\times 10^{-4}$.  A recent
re-observation of the high D/H absorption system has been resolved into 
two components, both yielding high values with an average value of D/H = $1.9
^{+0.4}_{-0.3} \times 10^{-4}$ \cite{rh1} as well as an
additional system with a similar high value \cite{rh2}. Other 
high D/H ratios were reported in \cite{quas3}. However, there are reported low
values of D/H in other such systems  \cite{quas2} with values D/H $\simeq 2.5
\times 10^{-5}$, significantly lower than the ones quoted above. It is probably
premature to use this value as the primordial D/H abundance in 
an analysis of big
bang nucleosynthesis, but it is certainly encouraging that 
future observations may
soon yield a firm value for D/H. It is however important to 
note that there does
seem to be a  trend that over the history of the Galaxy, the D/H ratio  is
decreasing, something we expect from galactic chemical evolution.  
Of course the
total amount of deuterium astration that has occurred is still uncertain, and
model dependent.

There are also several types of \he3 measurements. As noted above, meteoritic
extractions yield a presolar value for \he3/H as given in Eq. (\ref{he3}).
In addition, there are several ISM measurements of \he3 in galactic HII
regions \cite{bbbrw} which also show a wide dispersion
\beq
 {\rm \left({~^3He \over H} \right)_{HII}} \simeq 1 - 5 \times 10^{-5}
\eeq
There is also a recent ISM measurement of \he3 \cite{gg}
with
\beq
 {\rm \left({~^3He \over H} \right)_{ISM}} = 2.1^{+.9}_{-.8} \times 10^{-5}
\eeq
  Finally there are observations of \he3 in planetary
nebulae \cite{rood} which show a very high \he3 abundance of 
\he3/H $\sim 10^{-3}$.

Each of the light element isotopes can be made consistent with theory for a
specific range in $\eta$. Overall consistency of course requires that
the range in $\eta$ agree among all four light elements.
\he3 (together with D) has stood out in its importance for BBN, because 
it  provided a (relatively large) lower limit for the baryon-to-photon
ratio \cite{ytsso}, $\eta_{10} > 2.8$. This limit for a long 
time was seen to be
essential because it provided the only means for bounding $\eta$ from below
and in effect allows one to set an upper limit on the number of neutrino
flavors \cite{ssg}, $N_\nu$, as well as other constraints on particle physics
properties. That is, the upper bound to $N_\nu$ 
is strongly dependent on the lower bound to
$\eta$.  This is easy to see: for lower $\eta$, the \he4 abundance drops,
allowing for a larger $N_\nu$, which would raise the \he4 abundance.
However, for $\eta < 4 \times 10^{-11}$, corresponding to $\Omega h^2 \sim
.001-.002$, which is not too different from galactic mass densities, 
there is no
bound whatsoever on $N_\nu$ \cite{ossty}. Of course, with the improved data on
\li7, we do have lower bounds on $\eta$ which exceed $10^{-10}$.

 In \cite{ytsso}, it was argued that since stars (even massive stars) do not 
destroy \he3 in its entirety, we can obtain a bound on $\eta$ from an
upper bound to the solar D and \he3 abundances. One can in fact limit
\cite{ytsso,ped}
 the sum of primordial D and \he3 by applying the expression below at $t =
\odot$
\beq
{\rm \left({D + \he3 \over H} \right)_p \le \left({D \over H} \right)_t}
+ {1 \over g_3}{\rm  \left({\he3 \over H} \right)_t}
\label{he3lim}
\eeq
In (\ref{he3lim}), $g_3$ is the fraction of a star's initial D and \he3 which
survives as \he3. For $g_3 > 0.25$ as suggested by stellar models, 
and using the
solar data on D/H and
\he3/H, one finds $\eta_{10} > 2.8$.

The limit $\eta_{10} > 2.8$ derived using (\ref{he3lim}) is really a one
shot approximation.  Namely, it is assumed that material passes 
through a star no
more than once. To determine whether or not the solar (and present) 
values of D/H
and
\he3/H can be matched it is necessary to consider models of galactic chemical
evolution \cite{bt}. In the absence of stellar \he3 production, 
particularly by
low mass stars, it was shown \cite{vop} that there are indeed suitable choices
for a star formation rate and an initial mass function to: 1) match the D/H
evolution from a primordial value (D/H)$_p = 7.5 \times 10^{-5}$,
corresponding to $\eta_{10} = 3$, through the solar and ISM abundances, 
while 2)
at the same time keeping the \he3/H evolution relatively flat so as not to
overproduce \he3 at the solar and present epochs. This was achieved for $g_3 >
0.3$. Even for $g_3 \sim 0.7$, the present \he3/H could be matched, though the
solar value was found to be a factor of 2 too high. For (D/H)$_p 
\simeq 2 \times
10^{-4}$, corresponding to  $\eta_{10} \simeq 1.7$, models could be
found which destroy D sufficiently; however, overproduction of 
\he3 occurred unless
$g_3$ was tuned down to about 0.1.

In the context of models of galactic chemical evolution, there is, however, 
 little justification a
priori for neglecting the production of \he3 in low mass
stars. Indeed, stellar models predict that considerable amounts of \he3 are
produced in stars between 1 and 3 M$_\odot$. For M $<$ 8M$_\odot$, Iben and
Truran \cite{it} calculate
\beq
(^3{\rm He/H})_f = 1.8 \times 10^{-4}\left({M_\odot \over M}\right)^2 
+ 0.7\left[({\rm D+~^3He)/H}\right]_i
\label{it}
\eeq
so that at $\eta_{10} = 3$, and ((D + \he3)/H)$_i = 9 \times 10^{-5}$,
$g_3(1 $M$_\odot$) = 2.7! It should be emphasized that this prediction is in
fact consistent with the observation of high \he3/H in planetary nebulae
\cite{rood}.

Generally, implementation of the \he3 yield in Eq. (\ref{it}) in chemical
evolution models leads to an overproduction of \he3/H particularly at the
solar epoch \cite{orstv,galli}. In Figure 1, the evolution of D/H and 
\he3/H is
shown as a function of time from \cite{scostv,orstv}. The solid curves 
show the
evolution in a  simple model of galactic chemical evolution with a star 
formation
rate proportional to the gas density and a power law IMF (see \cite{orstv}) for
details). The model was chosen to fit the observed deuterium 
abundances. However,
as one can plainly see, \he3 is grossly overproduced (the deuterium data is
represented by squares and \he3 by circles). Depending on the particular model
chosen,  it may be possible to come close to at least the upper 
end of the range of
the \he3/H observed in galactic HII regions \cite{bbbrw}, 
although the solar value
is missed by many standard deviations.

\begin{figure}
\hspace{4truecm}
\epsfysize=20truecm\epsfbox{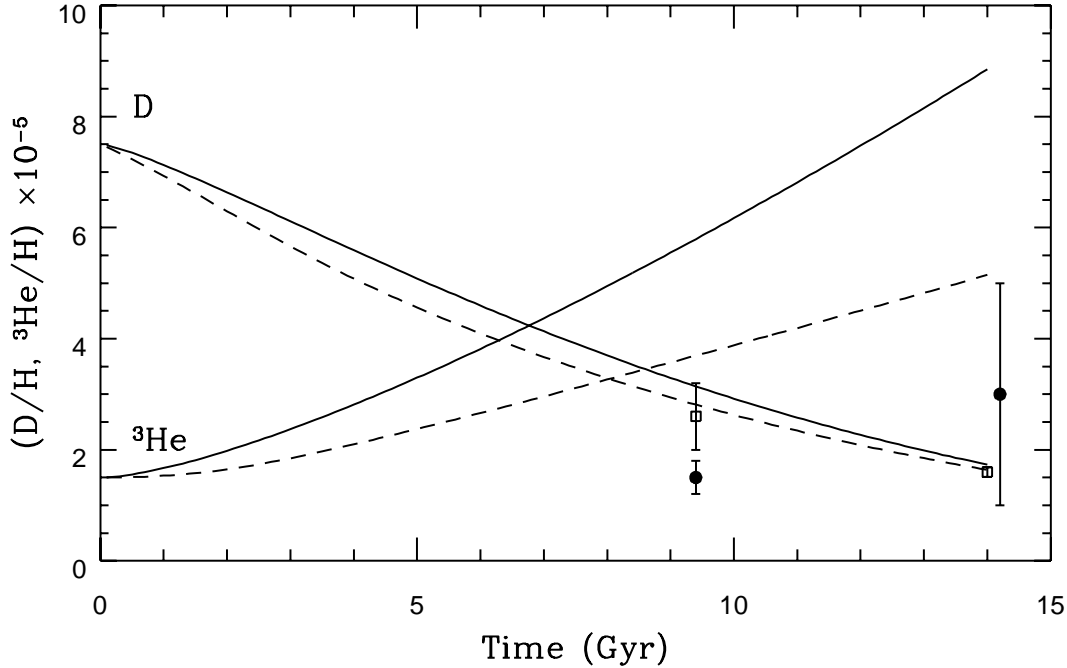}
\vspace{-10.5truecm}
\baselineskip=2ex
\caption { The evolution of D and \he3 with time.}
\end{figure}

\baselineskip=3ex
The overproduction of \he3 relative to the solar meteoritic value seems to be a
generic feature of chemical evolution models when \he3 production in low mass
stars is included. In \cite{scostv}, a more extreme model of galactic chemical
evolution was tested.  There, it was assumed that the initial mass function
was time dependent in such a way so as to favor massive stars early on (during
the first two Gyr of the galaxy).  Massive stars are preferential from the
point of view of destroying \he3.  However, massive stars are also proficient
producers of heavy elements and in order to keep the metallicity of the disk
down to acceptable levels, supernovae driven outflow was also included.
The degree of outflow was limited roughly by the observed metallicity in the
intergalactic gas in clusters of galaxies. One further assumption was 
necessary;
we allowed the massive stars to lose their \he3 depleted hydrogen
envelopes prior
to explosion.  Thus only the heavier elements were expulsed from the galaxy.
With all of these (semi-defensible) assumptions, \he3 was still overproduced
at the solar epoch, as shown by the dashed curve in Figure 1. Though there
certainly is an improvement in the evolution of \he3 without reducing the
yields of low mass stars, it is hard to envision much further reduction in
the solar \he3 predicted by these models. The
only conclusion that we can make at this point is that there is 
clearly something
wrong with our understanding of
\he3 in terms of either chemical evolution, stellar evolution or 
perhaps even the
observational data.

Given the magnitude of the problems concerning \he3, it would seem unwise to
make any strong conclusion regarding big bang nucleosynthesis 
which is based on
\he3.  Perhaps as well some caution is deserved with regard to the recent D/H
measurements, although if the present trend continues and is 
verified in several
different quasar absorption systems, then D/H will certainly become our best
measure for the baryon-to-photon ratio. Given the current situation 
however, it
makes sense to take a step back and perform an analysis of big bang
nucleosynthesis in terms of the element isotopes that are best understood,
namely \he4 and \li7.

Monte Carlo techniques are proving to be a useful form of analysis for big
bang nucleosynthesis \cite{kr,hata1}. In \cite{fo}, we performed just such an
analysis using only \he4 and \li7. It should be noted that in principle, two
elements should be sufficient for constraining the one parameter 
($\eta$) theory
of BBN. We begin by establishing likelihood functions for the theory and
observations. For example, for \he4, the theoretical likelihood 
function takes the
form
\beq
L_{\rm BBN}(Y,Y_{\rm BBN}) 
  = e^{-\left(Y-Y_{\rm BBN}\left(\eta\right)\right)^2/2\sigma_1^2}
\label{gau}
\eeq
where $Y_{\rm BBN}(\eta)$ is the central value for the \he4 mass fraction
produced in the big bang as predicted by the theory at a given value of $\eta$,
and $\sigma_1$ is the uncertainty in that  value derived from the Monte Carlo
calculations \cite{hata1} and is a measure of the theoretical 
uncertainty in the
big bang calculation. Similarly one can write down an expression for the
observational likelihood function. In this case we have two 
 sources of  errors, as
discussed above, a statistical uncertainty, $\sigma_2$ and a systematic
uncertainty,
$\sigma_{\rm sys}$.  Here, I will assume that the
systematic error is described by a top hat distribution \cite{hata1,osb}.
The convolution of the top hat distribution and the Gaussian (to describe
the statistical errors in the observations) results in the difference
of two error functions
\beq
L_{\rm O}(Y,Y_{\rm O}) = 
{\rm erf}\left({Y - Y_{\rm O} + \sigma_{\rm sys}  
     \over \sqrt{2} \sigma_2}\right) - 
{\rm erf}\left({Y - Y_{\rm O} - \sigma_{\rm sys} 
     \over \sqrt{2} \sigma_2}\right)
\label{erf}
\eeq
where in this case, $Y_{\rm O}$ is the observed 
(or observationally determined)
value for the \he4 mass fraction. (Had I used a Gaussian to describe the
systematic uncertainty, the convolution of two Gaussians leads to a
Gaussian, and the likelihood function (\ref{erf}) would have 
taken a form similar
to that in (\ref{gau}).

A total likelihood 
function for each value of $\eta_{10}$ is derived by
convolving the theoretical
and observational distributions, which for \he4 is given by
\beq
{L^{^4{\rm He}}}_{\rm total}(\eta) = 
\int dY L_{\rm BBN}\left(Y,Y_{\rm BBN}\left(\eta\right)\right) 
L_{\rm O}(Y,Y_{\rm O})
\label{conv}
\eeq
An analogous calculation is performed \cite{fo} for \li7. 
The resulting likelihood
functions from the observed abundances given in Eqs. (\ref{he4}) 
  and (\ref{li})
is shown in Figure 2. As one can see 
there is very good agreement between \he4 and \li7 in the vicinity
of $\eta_{10} \simeq 1.8$.

\begin{figure}
\hspace{0.5truecm}
\epsfysize=9truecm\epsfbox{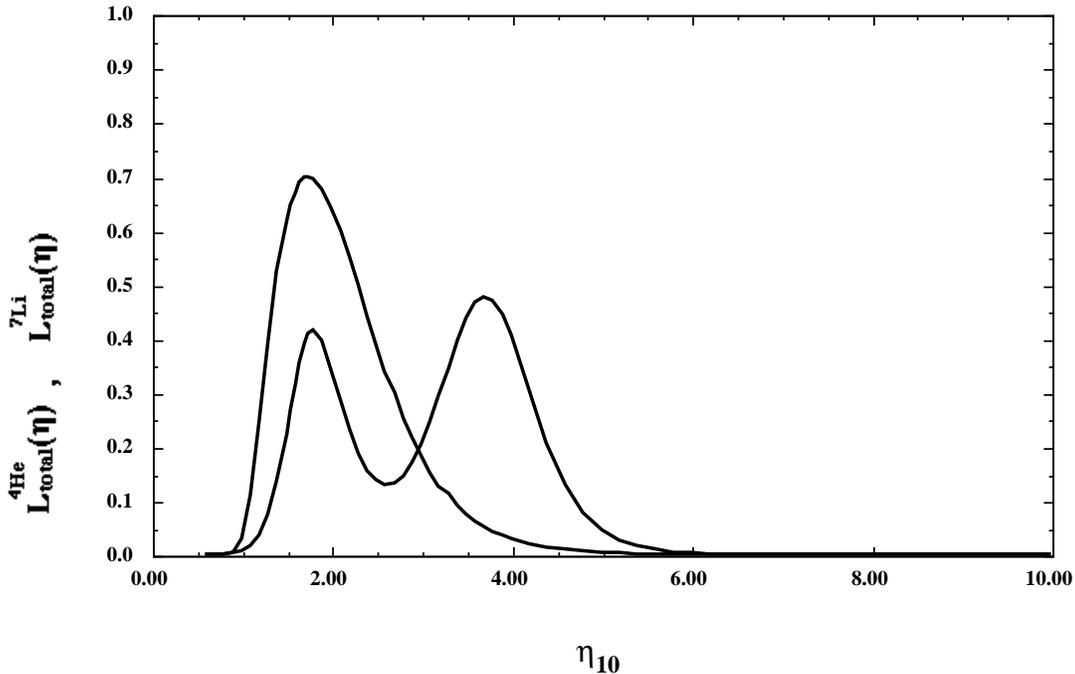}

\caption { \baselineskip=2ex Likelihood distribution for each of \he4 and \li7,
shown as a  function of $\eta$.  The one-peak structure of the \he4 curve
corresponds to its monotonic increase with $\eta$, while
the two-peaks for \li7 arise from its passing through a minimum.}
\label{fig:fig1}
\end{figure}

\begin{figure}
\hspace{0.5truecm}
\epsfysize=9truecm\epsfbox{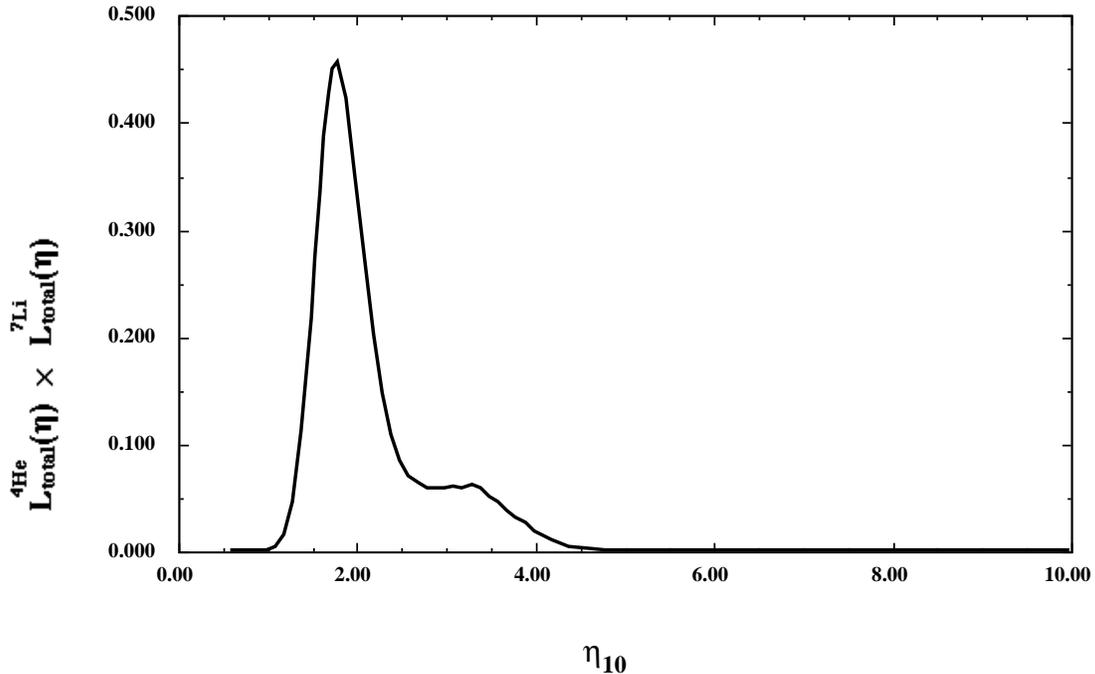}
\baselineskip=2ex
\caption { Combined likelihood for simultaneously fitting \he4 and \li7,
as a function of $\eta$.
}
\label{fig:fig2}
\end{figure}

\baselineskip=3ex

The combined likelihood, for fitting both elements simultaneously,
is given by the product of the two functions in Figure \ref{fig:fig1}
and is shown  in Figure \ref{fig:fig2}.
{}From Figure \ref{fig:fig1} it is clear that \he4 overlaps
the lower (in $\eta$) \li7 peak, and so one expects that 
there will be concordance
in an allowed range of $\eta$ given by the overlap region.  
This is what one finds in Figure \ref{fig:fig2}, which does
show concordance and gives a preferred value for $\eta$, 
$\eta_{10}  = 1.8^{+1}_{-.2}$ corresponding to $\Omega_B h^2 =
.006^{+.004}_{-.001}$.  

Thus,  we can conclude that 
the abundances of 
\he4 and \li7 are consistent, and select an $\eta_{10}$ range which
overlaps with (at the 95\% CL) the longstanding favorite
 range around $\eta_{10} = 3$.
Furthermore, by finding concordance  
using only \he4 and \li7, we deduce that
if there is problem with BBN, it must arise from 
D and \he3 and is thus tied to chemical evolution or the stellar evolution of
\he3. The most model-independent conclusion is that standard
BBN  with $N_\nu = 3$ is not in jeopardy, 
but there may be problems with our
detailed understanding of D and particularly \he3
chemical evolution. It is interesting to
note that the central (and strongly)  peaked
value of $\eta_{10}$ determined from the combined \he4 and\li7 likelihoods
is at $\eta_{10} = 1.8$.  The corresponding value of D/H is 1.8 $\times 
10^{-4}$, very close to the high value  of D/H in quasar absorbers
\cite{quas1,rh1,quas3}.
Since  D and \he3 are monotonic functions of $\eta$, a prediction for 
$\eta$, based on \he4 and \li7, can be turned into a prediction for
D and \he3.  
 The corresponding 95\% CL ranges are D/H  $= (5.5 - 27)  \times
10^{-5}$ and  and \he3/H $= (1.4 - 2.7)  \times 10^{-5}$.

If we did have full confidence in the measured value of D/H in 
quasar absorption
systems, then we could perform the same statistical analysis 
using \he4, \li7, and
D. To include D/H, one would
proceed in much the same way as with the other two light elements.  We
compute likelihood functions for the BBN predictions as in
Eq. (\ref{gau}) and the likelihood function for the observations using
D/H = $(1.9 \pm 0.4) \times 10^{-4}$.  We are using only the high
 value of D/H
here. These are
then convolved as in Eq.  (\ref{conv}).  
In figure 4, the resulting normalized
distribution, $L^{{\rm D}}_{\rm total}(\eta)$ is super-imposed on
distributions appearing in figure 2. 
It is indeed startling how the three peaks, for
D, \he4 and \li7 are literally on top of each other.  In figure 5, 
the combined distribution is shown.
We now  have a very clean distribution and prediction 
for $\eta$, $\eta_{10}  = 1.75^{+.3}_{-.1}$ corresponding to $\Omega_B h^2 =
.006^{+.001}_{-.0004}$,
with the peak of the distribution at $\eta_{10} = 1.75$.  
The absence of any overlap with the high-$\eta$ peak of the \li7
distribution has considerably lowered the upper limit to $\eta$. 
Overall, the concordance limits in this case are dominated by the 
deuterium likelihood function.

\begin{figure}
\hspace{0.5truecm}
\epsfysize=9truecm\epsfbox{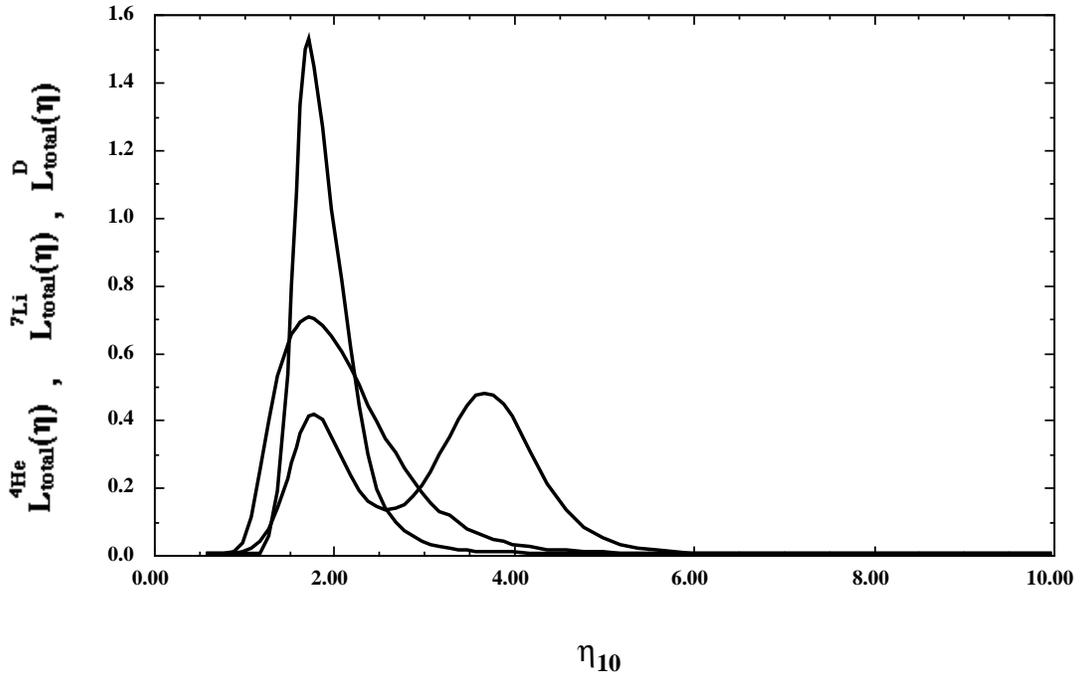}
\baselineskip=2ex
\caption { As in Figure 2, with the addition of the likelihood 
distribution for
D/H. }
\label{fig:fig4}
\end{figure}

\baselineskip=3ex

\begin{figure}
\hspace{0.5truecm}
\epsfysize=9truecm\epsfbox{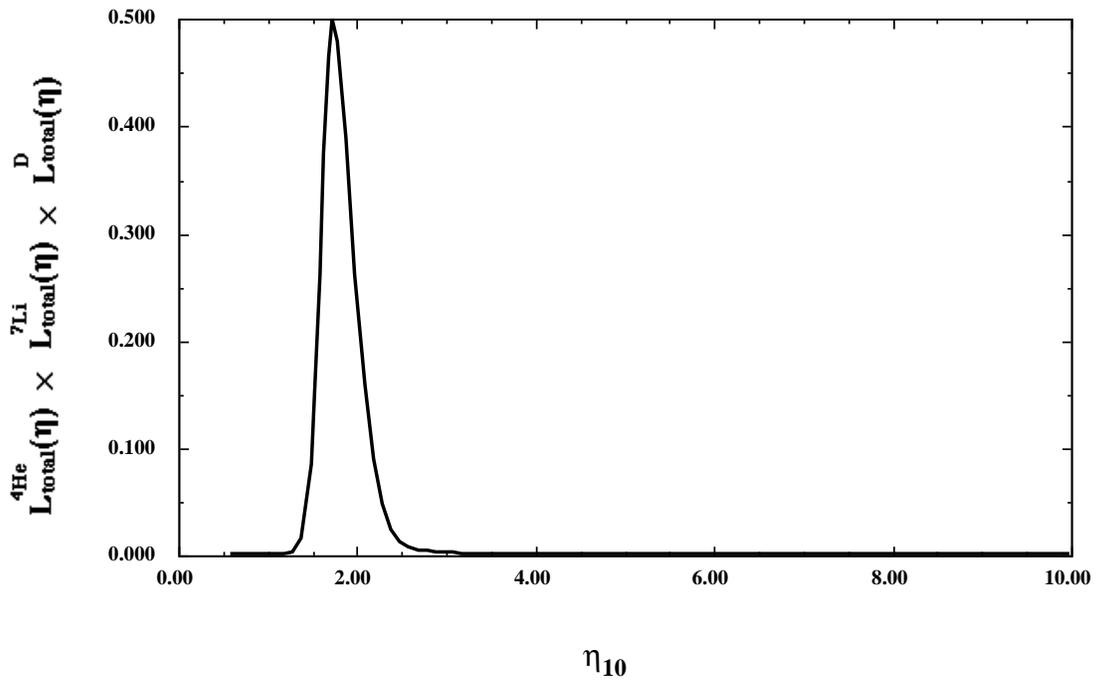}
\baselineskip=2ex
\caption { Combined likelihood for simultaneously fitting 
\he4 and \li7, and D 
as a function of $\eta$.
}
\label{fig:fig5}
\end{figure}

\baselineskip=3ex

To summarize on the subject of big bang nucleosynthesis, 
I would assert that one
can conclude that the present data on the abundances of the light element
isotopes are consistent with the standard model of big bang 
nucleosynthesis. Using
the the isotopes with the best data, \he4 and
\li7, it is possible to constrain the theory and obtain a best value for the
baryon-to-photon ratio of $\eta_{10} = 1.8$, a corresponding 
value $\Omega_B =
.0065$ and
\begin{eqnarray}
1.4 & < \eta_{10} & < 3.8  \qquad 95\% {\rm CL} \nonumber \\
.005 & < \Omega_B h^2 & < .014  \qquad 95\% {\rm CL}
\label{res2}
\end{eqnarray}
For $0.4 < h < 1$, we have a range $ .005 < \Omega_B < .09$.
This is a rather low value for the baryon density
 and would suggest that much of the galactic dark matter is
non-baryonic \cite{vc}. These predictions are in addition 
consistent with recent
observations of D/H in quasar absorption systems which show a high value.
Difficulty remains however, in matching the solar \he3 abundance, suggesting a
problem with our current understanding of galactic chemical evolution or the
stellar evolution of low mass stars as they pertain to \he3.

If we now take as given that non-baryonic dark matter is required, 
we are faced
with the problem of its identity.
	Light neutrinos ($m \le 30 eV$) are 
a long-time standard when it comes to
 non-baryonic dark matter \cite{ss}.  Light neutrinos produce 
structure on large scales and the natural (minimal) scale for
 structure clustering is given in Eq. (\ref{mj}).  Hence neutrinos
 offer the natural possibility for large scale structures \cite{nu1,nu2} 
including filaments and voids.  It seemed, however, that neutrinos
 were ruled out because they tend to
 produce too much large scale structure \cite{nu3}.
  Because the smallest non-linear structures have mass scale $M_J$ and 
the typical galactic mass scale is $\simeq 10^{12} M_\odot$, galaxies must 
fragment out of the larger pancake-like objects.  The problem is 
that in such a scenario, galaxies form late \cite{nu2,nu4} 
 ($z \le 1$) whereas
 quasars and galaxies are seen out to redshifts $z \ga 4$. 
Recently, neutrinos are seeing somewhat of a revival 
in popularity in mixed dark matter models.

In the standard model, the absence of a right-handed neutrino state precludes
the existence of a neutrino mass.  By adding a right-handed 
state $\nu_R$, it is
possible to generate a Dirac mass for the neutrino,
 $m_\nu = h_\nu v/\sqrt{2}$,
as for the charged lepton masses, where $h_\nu$ is the neutrino Yukawa coupling
constant, and $v$ is the Higgs expectation value.  It is also possible to
generate a Majorana mass for the neutrino when in addition to the Dirac mass
term, $m_\nu \bar{\nu_R} \nu_L$, a term $M \nu_R \nu_R$ is included.
In what is known as the see-saw mechanism, the two mass eigenstates are given
by $m_{\nu_1} \sim m_\nu^2/M$ which is very light, and $m_{\nu_2}
 \sim M$ which is
heavy.  The state $\nu_1$ is our hot dark matter candidate as 
$\nu_2$ is in general
not stable.

	The cosmological constraint on the mass of a
 light neutrino is derived from the overall mass density of 
the Universe.  In general, the mass density of a light particle $\chi$ can be
expressed as
\beq
	\rho_\chi  = m_\chi Y_\chi n_\gamma  \le \rho_c  = 
1.06 \times 10^{-5} {h_o}^2 {\rm GeV/cm}^3	
\eeq
where $Y_\chi = n_\chi/n_\gamma$ is the density of $\chi$'s relative
 to the density of
 photons, for $\Omega {h_o}^2 < 1$.  For neutrinos $Y_\nu = 3/11$,
 and one finds \cite{cows}
\beq
	\sum_\nu (\frac{g_\nu}{2}) m_\nu  < 93 {\rm eV} (\Omega {h_o}^2)	
\label{ml}
\eeq
where the sum runs over neutrino flavors.  All particles with 
abundances $Y$ similar to neutrinos will have a mass limit 
given in Eq. (\ref{ml}).

 It was  possible that neutrinos (though not any of the 
known flavors) could have had large masses, $m_\nu > 1$ MeV.  In 
that case their abundance $Y$ is controlled by $\nu,{\bar \nu} $
 annihilations \cite{lw}, 
for example, $\nu {\bar \nu} \rightarrow f {\bar f}$ via Z exchange.  
When the annihilations 
freeze-out (the annihilation rate becomes slower than the expansion rate 
of the Universe), $Y$ becomes fixed.  Roughly, $Y \sim (m\sigma_A)^{-1}$
 and $\rho \sim {\sigma_A}^{-1}$
 where $\sigma_A$ is the annihilation cross-section.  For neutrinos, 
we expect $\sigma_A \sim {m_\nu}^2/{m_Z}^4$ so that $\rho_\nu 
\sim 1/{m_\nu}^2$  and
 we can derive a lower bound \cite{ko,wso} on the neutrino mass,
 $m_\nu \ga 3-7$ GeV, depending on whether it is a Dirac 
or Majorana neutrino. Indeed, any particle with roughly a weak scale
cross-sections will tend to give an interesting value of $\Omega h^2 \sim 1$
\cite{jung}.

  Due primarily to the limits from LEP \cite{lep}, the heavy 
massive neutrino has become simply an example and is no longer a
 dark matter candidate. LEP excludes neutrinos (with standard weak
interactions) with masses $m_\nu \la 40$ GeV.  Lab constraints for 
Dirac neutrinos are available \cite{dir}, excluding neutrinos 
with masses between
10 GeV and 4.7 TeV. This is significant, since it precludes the possibility 
of neutrino dark matter based on an asymmetry between $\nu$ and ${\bar \nu}$ 
\cite{ho}. Majorana neutrinos are excluded as {\em dark matter}
since $\Omega_\nu {h_o}^2 < 0.001$ for $m_\nu > 40$ GeV and are thus
cosmologically  uninteresting.

Supersymmetric theories introduce several possible candidates. 
If R-parity, which distinguishes between ``normal" matter and the 
supersymmetric partners and can be defined in terms of baryon, lepton and
spin as $R = (-1)^{3B + L + 2S}$, is unbroken, there is at least one 
supersymmetric particle which must be stable.  I will assume R-parity
conservation, which is common in the MSSM. R-parity is generally
 assumed in order
to justify the absence of superpotential terms can be responsible for rampid
proton decay. 
 The stable 
particle (usually called the LSP) is most probably some 
linear combination of the only $R=-1$ neutral fermions, the 
neutralinos \cite{ehnos}: the wino $\tilde W^3$, the partner of the
 3rd component of the $SU(2)_L$ gauge boson;
 the bino, $\tilde B$, the partner of the $U(1)_Y$ gauge boson;
 and the two neutral Higgsinos,  $\tilde H_1$ and $\tilde H_2$.
 Gluinos are expected to be heavier -$m_{\tilde g} = (\frac{\alpha_3}{\alpha})
 \sin^2 \theta_W M_2$, where $M_2$ is the supersymmetry breaking SU(2) gaugino
mass-  and they do not mix with the other states.
  The sneutrino \cite{snu} is also a possibility but has been
 excluded as a dark 
matter candidate by direct \cite{dir} searches, indirect \cite{indir}
 and accelerator\cite{lep} 
searches.  For more on the motivations for supersymmetry and 
the supersymmetric
parameter space, see the contribution of Jungman \cite{jung}.

The identity of the LSP is effectively determined by three 
parameter in the MSSM,
the gaugino mass, $M_2$, the Higgs mixing mass $\mu$, and the 
ratio of the Higgs
vacuum expectation values, $\tan \beta$. 	
In Figure 6 \cite{osi3}, regions in
the $M_2, \mu$
 plane with $\tan\beta = 2$ are shown in which the LSP
 is one of several nearly pure states, the photino, $\tilde \gamma$, the U(1)
gaugino,
$\tilde B$, a symmetric combination of the Higgsinos, 
$\tilde{H}_{(12)} = {1 \over
\sqrt{2}}
  ( {{\tilde H}_1} +  {{\tilde H}_2})$, or the Higgsino 
$\tilde{S} = \tilde H_1
\cos\beta + \tilde H_2 \sin\beta$. The dashed lines show the LSP mass contours.
 The cross hatched regions correspond to parameters giving
  a chargino ($\tilde W^{\pm}, \tilde H^{\pm}$) state 
with mass $m_{\tilde \chi} \leq 45 GeV$ and as such are 
excluded by LEP\cite{lep2}.
This constraint has been extended by LEP1.5, \cite{lep15,efos} 
and is shown by the 
light shaded region and corresponds to regions where the chargino mass is $\la
67$ GeV.  The dark shaded region corresponds to a limit on
$M_2$ from the limit\cite{cdf}
 on the gluino mass $m_{\tilde g} \leq 70$ GeV or $M_2 \leq 22$ GeV.
 Notice that the parameter space is dominated by the  
$\tilde B$ or $\tilde H_{12}$
 pure states and that the photino (most often quoted as the LSP)
 only occupies a small fraction of the parameter space,
 as does the Higgsino combination $\tilde S^0$.

\begin{figure}
\hspace{0.5truecm}
\epsfysize=11truecm\epsfbox{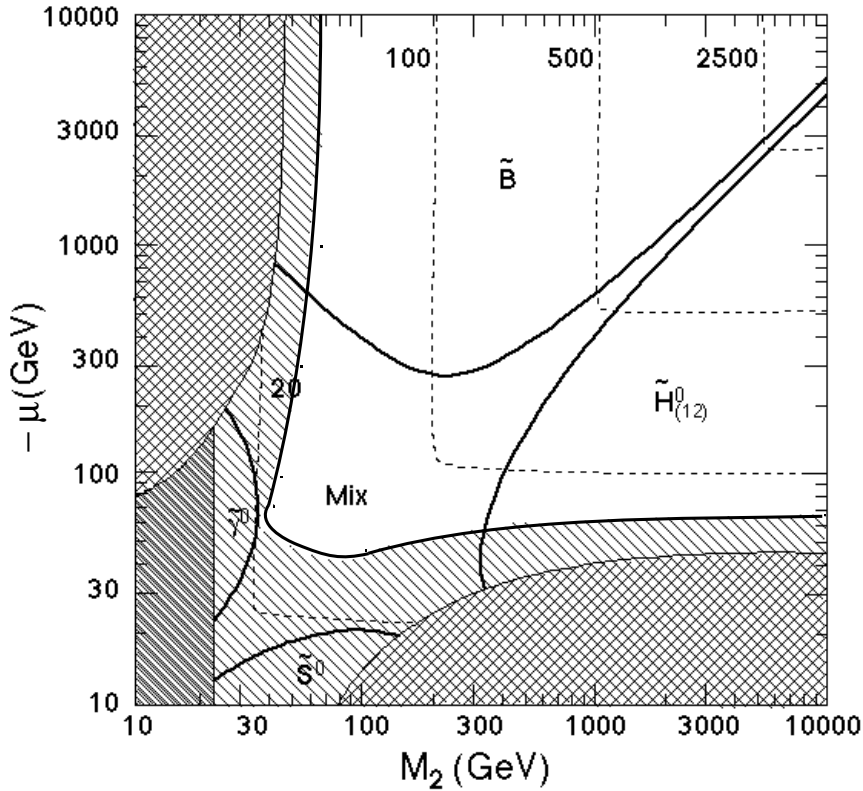}
\baselineskip=2ex
\caption { The $M_2$-$\mu$ plane in the MSSM for $\tan \beta = 2$.
}
\label{fig:fig6}
\end{figure}

\baselineskip=3ex

As described in \cite{jung}, the relic abundance of LSP's is 
determined by solving
the Boltzmann
 equation for the LSP number density in an expanding Universe.
 The technique\cite{wso} used is similar to that for computing
 the relic abundance of massive neutrinos\cite{lw}.
 For binos, as was the case for photinos \cite{phot}, it is possible
 to adjust the sfermion masses $m_{\tilde f}$ to obtain closure density.
Adjusting the sfermion mixing parameters \cite{fkmos} or CP violating phases
\cite{fkos} allows even greater freedom.
 In Figure 7 \cite{70}, the relic abundance ($\Omega h^2$) is shown in the
$M_2-\mu$
 plane with $\tan\beta = 2$, the Higgs pseudoscalar mass $m_0 = 50$ GeV,
 $m_t = 100$ GeV, and  $m_{\tilde f} = 200$ GeV.
 Clearly the MSSM offers sufficient room to solve the dark matter problem.
Similar results have been found by other groups \cite{gkt,dn,dvn}.
In Figure 7, in the higgsino sector ${\tilde H}_{12}$ marked off by the dashed
line,
 co-annihilations \cite{gs,dn}
 between ${\tilde H}_{(12)}$ and the next lightest 
neutralino (also a Higgsino)
were not included. These tend to lower significantly 
the relic abundance in much
of this sector.

\begin{figure}
\hspace{0.5truecm}
\epsfysize=11truecm\epsfbox{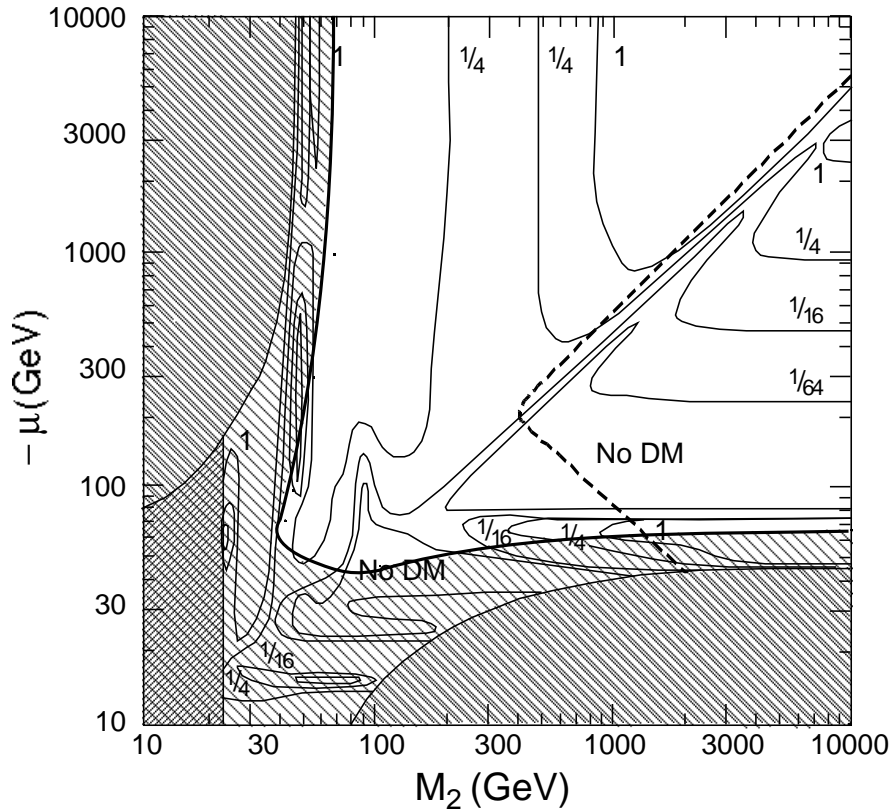}
\baselineskip=2ex
\caption { Relic density contours ($\Omega h^2$) in the $M_2$ - $\mu$ plane.
}
\label{fig:fig7}
\end{figure}

\baselineskip=3ex

Though I have concentrated on the LSP in the MSSM as a cold dark matter
candidate, there are many other possibilities when one goes 
beyond the standard
model.  Axions were discussed at length by Jungman \cite{jung} 
and Lu \cite{lu}.
A host of other possibilities were discussed by Khlopov \cite{khl}.

The final subject that I will cover in this introduction/summary 
is the question
of detection.  Dark matter detection can be separated into two basic methods,
direct \cite{gond} and indirect \cite{berg}.  Direct detection relies on the
ability to detect the elastic scattering of a dark matter 
candidate off a nucleus
in a detector.  The experimental signatures for direct 
detection were covered by
Cabrera \cite{cab} and several individual experiments were
 described \cite{expt}.

The detection rate, will depend on the density of dark matter in the solar
neighborhood, $\rho \sim 0.3$ GeV/cm$^3$, the velocity, $v \sim 300$ km/s, and
the elastic cross section, $\sigma$.  Spin independent 
interactions are the most
promising for detection.  Dirac neutrinos have spin-independent 
interactions, but
as noted above, these have already been excluded as dark matter by direct
detection experiments \cite{dir}. In the MSSM, it is possible for 
the LSP to also
have spin independent interactions which are mediated by 
Higgs exchange.  These
scatterings are only important when the LSP is a mixed 
(gaugino/Higgsino) state
as in the central regions of Figures 6 and 7. Generally, 
these regions have low
values of $\Omega h^2$ (since the annihilation cross sections
 are also enhanced)
and the parameter space in which the elastic cross section and 
relic density are
large is rather limited.  Furthermore, a significant detection rate 
in this case
relies on a low mass for the Higgs scalar \cite{bfg,fos}.

More typical of the SUSY parameter space is a LSP with spin dependent
interactions.  Elastic scatterings are primarily spin dependent 
whenever the LSP
is mostly either gaugino or Higgsino. For Higgsino dark matter, 
Higgsinos with
scatterings mediated by $Z^0$ avoid the
$\tilde{H}_{(12)}$ regions of Figures 6 and 7, and as such are now largely
excluded (the $S^0$ region does grow at low $\tan \beta$ \cite{ehnos,osi3}.
Higgsino scatterings mediated by sfermion exchange depend on couplings
proportional to the light quark masses and will have cross sections which are
suppressed by $(m_p/m_W)^4$, where $m_p$ is the proton mass. These rates are
generally very low \cite{fos}.  Binos, on the other hand, will 
have elastic cross
sections which go as $m^2/{m_{\tilde f}}^4$, where $m$ is the
 reduced mass of the
bino and nucleus. These rates are typically higher (reaching up to almost 0.1
events per kg-day \cite{fos,ef,bg}.

Indirect methods also offer the possibility for the detection of dark matter.
Three methods for indirect detection were discussed \cite{berg}.
 1) $\gamma$-rays
from dark matter annihilations in the galactic halo are a possible signature
\cite{gam}. In the case of the MSSM, unless the mass of the LSP
 is larger than
$m_W$, the rates are probably too small to be detectable over background
\cite{berg}.  2) Dark matter will be trapped gradually in the sun, and
annihilations within the sun will produce high energy neutrinos which may be
detected \cite{slkosi}; similarly, annihilations within the earth 
may provide a
detectable neutrino signal \cite{earth}. Edsjo \cite{edsjo} discussed
possibilities for determining the mass the dark matter candidate 
from the angular
distribution of neutrinos.  This method hold considerable promise, 
as there will
be a number of very large neutrino detectors coming on line in the future.
Finally, 3) there is the possibility that halo annihilations 
into positrons and
antiprotons in sufficient numbers to distinguish them from 
cosmic-ray backgrounds
\cite{gam,DeR,tar}.

\vskip 1in
\vbox{
\noindent{ {\bf Acknowledgments} } \\
\noindent    I would like to thank T. Falk for his help in proof reading
this manuscript. This work was supported in part by DOE grant
DE--FG02--94ER--40823.}


\end{document}